\documentclass[conference]{IEEEtran}
\IEEEoverridecommandlockouts

\usepackage{cite}
\usepackage{amsmath,amssymb,amsfonts}
\usepackage{algorithmic}
\usepackage{graphicx}
\usepackage{textcomp}
\usepackage{xcolor}
\usepackage{comment}
\def\BibTeX{{\rm B\kern-.05em{\sc i\kern-.025em b}\kern-.08em
    T\kern-.1667em\lower.7ex\hbox{E}\kern-.125emX}}

\newcommand*{\affmark}[1][*]{\textsuperscript{#1}}

\begin{document}

\title{\LARGE Blockage Prediction in Directional mmWave Links Using Liquid Time Constant Network
}

\author{\IEEEauthorblockN{Martin H. Nielsen\affmark[1,2], Chia-Yi Yeh\affmark[1], Ming Shen\affmark[2], and Muriel Médard\affmark[1]}
\IEEEauthorblockA{\affmark[1]\textit{Massachusetts Institute of Technology, Cambridge, MA, 02139 USA} \\
\affmark[2]\textit{Aalborg University, Aalborg, 9000 DK}} \\ \vspace{-1.2cm}} 

\maketitle

\begin{abstract}
We propose to use a liquid time constant (LTC) network to predict the future blockage status of a millimeter wave (mmWave) link using only the received signal power as the input to the system. 
The LTC network is based on an ordinary differential equation (ODE) system inspired by biology and specialized for near-future prediction for time sequence observation as the input.
Using an experimental dataset at 60 GHz, we show that our proposed use of LTC can reliably predict the occurrence of blockage and the length of the blockage without the need for scenario-specific data. The results show that the proposed LTC can predict with upwards of 97.85\% accuracy without prior knowledge of the outdoor scenario or retraining/tuning. These results highlight the promising gains of using LTC networks to predict time series-dependent signals, which can lead to more reliable and low-latency communication.
\end{abstract}


\section{Introduction}
\IEEEPARstart{M}{illimeter} (mmWave) and terahertz (THz) communication is a promising technology for achieving high data rates in 6G and beyond. However, highly directional transmission in these bands makes the link vulnerable to blockage, causing sudden interruptions to a communication link.
Consequently, reliable mmWave and THz communication requires predicting the occurrence of blockage \cite{b2}.

Blockage prediction using machine learning based approaches has been investigated in the literature for both in-band and out-of-band solutions. While the state-of-the-art methods rely on information from multiple sources, either from other frequency spectrums \cite{b3} or a camera feed \cite{b4}, it is more advantageous if we can predict without these additional resources or hardware. Prior work also proposed to use prior statistical observations for meta-learning \cite{b5}, however, it requires prior information about the surrounding and thus is costly to generalize. In \cite{b6}, an in-band proactive predictor based on a pre-blockage power signature was proposed using deep learning to enable the prediction of the wireless link status. However, the  prediction accuracy decreases significantly when predicting even just slightly further into the future.

In this work, we propose to use Liquid Time Constant (LTC) network \cite{b7} for blockage prediction. Since the LTC network is based on ordinary differential equations (ODEs), it has strong expressiveness, stability, and performance in handling time series. Using the publicly available dataset on mmWave blockage scenarios \cite{b6}, we demonstrate the superior generalization of LTC networks by training using the indoor dataset, which contains pre-blockage power signature under controlled blockage movement and apply the trained model to the outdoor dataset with uncontrolled blockage events. We show accuracy in blockage prediction above 97.85\% for all outdoor scenarios for the immediate future timeslot ($t+1$). In addition, we demonstrate a 12\% to 39\% accuracy improvement in predicting blockage in the near future ($t+5$ and $t+10$ timeslots) compared to the baseline \cite{b6}. The proposed model's sparse network and high generalization capabilities make it an attractive option for pre-blockage prediction in high-frequency communication systems.

\begin{figure}
	\includegraphics[width=0.45\textwidth]{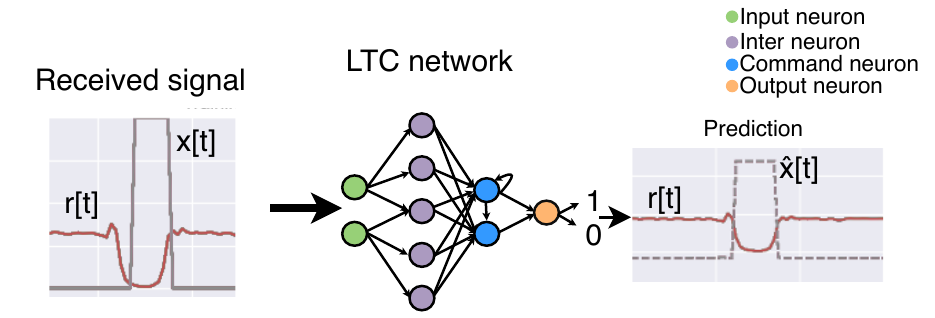}
    \caption{Proposed architecture and system model for THz blockage prediction $x[t]$ using LTC network. The input data is the received signal as a function of time samples $r[t]$. The LTC is trained to recognize the pre-blockage signature contained in the data. }
    \label{fig:System}
\end{figure}

\section{Problem formulation }

We address the problem of proactively identifying directional link blockage status using received mmWave signal power. 
We consider a system with one transmitter and one receiver, each equipped with a directional beam. After the link is established, the link can experience short blockages at unknown times.

We formulate the blockage prediction as a discrete-time problem where $t\ \in\mathbb{Z}$ is the index of the time samples. The received signal power is represented by $r[t]$, and the link blockage status is defined as $x[t] \in \{0,1\}$ where 1 indicates blockage and 0 indicates no blockage. 
At time $t$, given the received signal power samples $S_{ob}$ with an observation duration of $T_{ob}$, we predict whether the link is blocked in the future $T_k$ timeslots. That is,
\begin{equation}
    S_{ob}=\{ r[t-T_{ob}+1], \cdots, r[t-1],r[t]\},
\end{equation}
and the prediction is represented by 
\begin{equation}
    \hat{x}[t+K], \forall K \in \{1, \cdots, T_k\}
\end{equation}
Our goal is to maximize the blockage prediction success probability: 
\begin{equation}
    \max \mathbb{P}(\hat{x}[t+K] = x[t+K]  |S_{ob}), \; \forall K \in \{1, \cdots, T_k\}.
\end{equation}
In this paper, we propose to solve the blockage prediction problem using a LTC network, as illustrated in Fig. \ref{fig:System}. 


\section{Proposed Architecture with Liquid Time Constant Network}

Liquid Time Constant (LTC) network was first proposed in \cite{b7}, where LTC is introduced as variations of continuous time models loosely inspired by biological signals.
Since LTC networks are based on a system of ordinary differential equations (ODEs), they have strong expressiveness, stability, and performance in modeling time series over a short to medium time.
For time-series prediction tasks, the LTC network has been shown to outperform other modern RNNs, and long short-term memory networks on most metrics \cite{b7}. Moreover, the LTC network requires fewer neurons and generalizes easier than conventional RNN networks \cite{b7}.
This behavior of the LTC networks makes it a perfect fit for handling the time series blockage prediction problem.

Neural Circuit Policy (NCP), which creates a sparse network according to \cite{b8}, is jointly employed to implement an LTC network with fewer resources. Further, the NCPs allow the LTC network to be adaptable to previously unseen data and provide robustness to nonideal data samples, which enables offline training using pre-blockage signature. Once the LTC network is trained, it is deployed to make blockage prediction at each time slot $t$ for the future $T_k$ time slots.

\section{Evaluation}

To evaluate the performance of an LTC network on blockage prediction, we use a publicly available dataset collected with a directional transmitter and a directional receiver at 60 GHz, consisting of indoor and outdoor scenarios \cite{b3,b6}. For the indoor scenario, a controlled blockage event happened during data collection. In comparison, for the outdoor scenarios, uncontrolled blockage events were caused by passing vehicles.
The dataset includes the power readings and the ground truth labels for blockage for all time instances.
We use the pre-blockage signature under a controlled blockage \cite{b6} from the indoor dataset to train the neural network. We then use the trained LTC network to predict blockage events on the outdoor dataset.

We construct an 8-neuron LTC network consisting of two input neurons, four inter neurons, two control neurons, and one output neuron as described in \cite{b7} and illustrated in Fig. \ref{fig:System}.
We then use the NCPs defined in \cite{b8} to construct the neural network, which helps create a very sparse and efficient network for training. 
We define the input as a series of time data with 64 beam indexes, with power values over time. The output is then determined by the output neuron, either 0 or 1.  We then train the neural network for 40 epochs over the complete indoor data set using an Adam optimizer, a cross-entropy loss function, and a learning rate of 0.02. 
After the training, the LTC network is evaluated on the outdoor scenarios. Since not all 64 beams receive a strong enough signal, we exclude the predictions where the received power is lower than 0.4 normalized received power. 




\begin{figure}[ht]
	\includegraphics[width=0.45\textwidth]{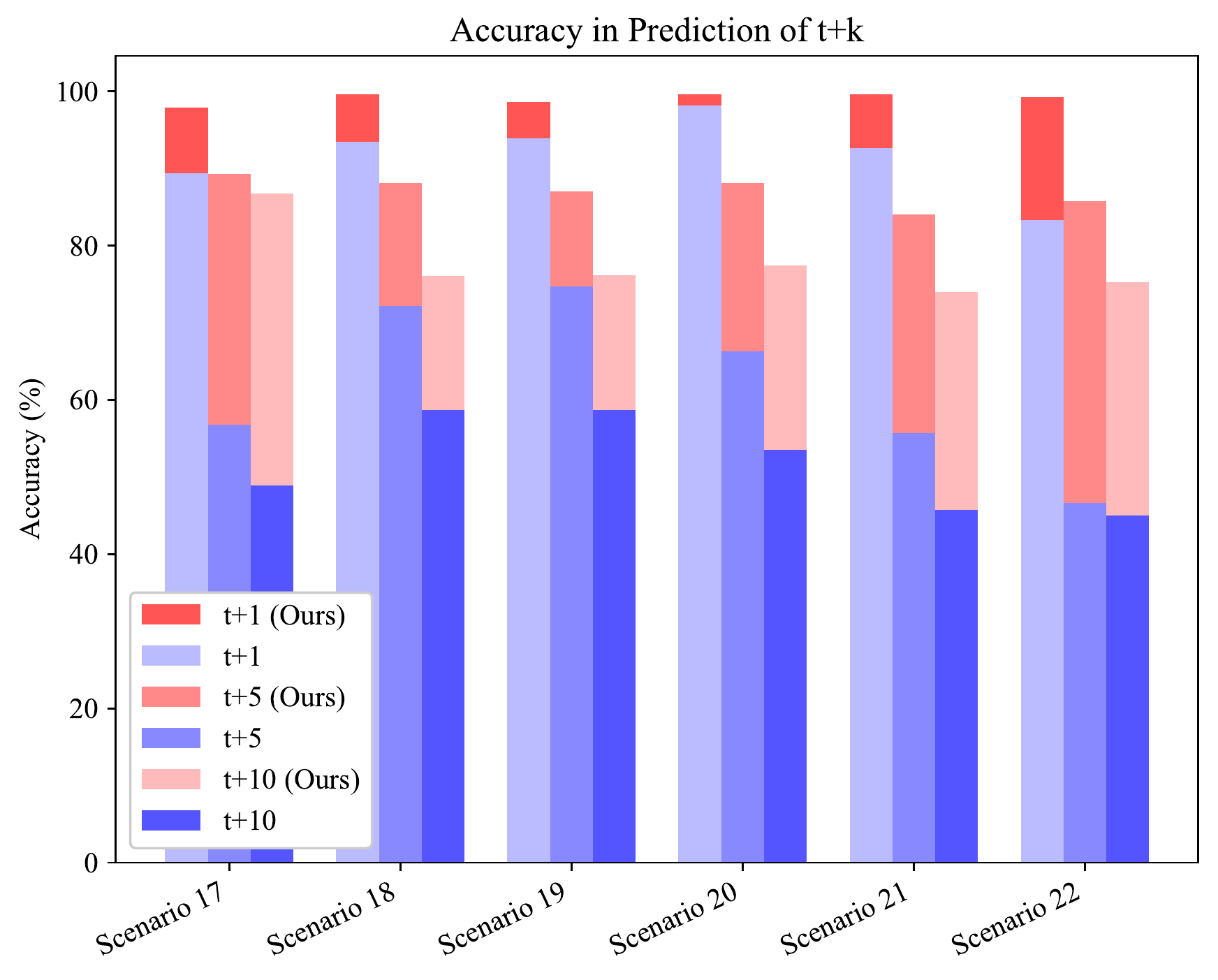}
    \caption{Blckage prediction accuracy of different outdoor scenarios, using our approach (red) and the baseline approach \cite{b3} (blue).}
    \label{fig:Results2}
\end{figure}


Fig. \ref{fig:Results2} presents the accuracy of blockage prediction in different outdoor scenarios with the scenario numbers defined in the dataset.
Fig. \ref{fig:Results2} shows that our proposed model outperforms the baseline approach given in \cite{b3}. The LTC network accuracy scores range from 73.95\% to 99.6\%, with the lowest score being significantly better than the state-of-the-art. Our approach shows, in some scenarios, an above 80\% accuracy for  $t+5$ and above 70\% for $t+10$, indicating our model's advantage in predicting further time in the future. The proposed model's sparse network and high generalization capabilities make it an attractive option for pre-blockage prediction in sub-terahertz communication systems. 

Further, the neural network can be quickly deployed and does not need to be trained for individual environments but shows high generalization capabilities for pre-blockage prediction. For future work, we will explore how this pre-blockage prediction can help improve reliability and latency in THz communication systems.

\begin{thebibliography}{00}
\bibitem{b2}Akyildiz, Ian, et al. "TeraHertz Band Communication: An Old Problem Revisited and Research Directions for the Next Decade." arXiv, 2021, https://doi.org/10.48550/arXiv.2112.13187.
\bibitem{b3}M. Alrabeiah and A. Alkhateeb, "Deep Learning for mmWave Beam and Blockage Prediction Using Sub-6 GHz Channels," in IEEE Transactions on Communications, vol. 68, no. 9, pp. 5504-5518, Sept. 2020, doi: 10.1109/TCOMM.2020.3003670.
\bibitem{b4}Alkhateeb, Ahmed, and Beltagy, Iz. "Machine Learning for Reliable mmWave Systems: Blockage Prediction and Proactive Handoff." arXiv, 2018,  https://doi.org/10.48550/arXiv.1807.02723.
\bibitem{b5}A. E. Kalør, O. Simeone and P. Popovski, "Prediction of mmWave/THz Link Blockages Through Meta-Learning and Recurrent Neural Networks," in IEEE Wireless Communications Letters, vol. 10, no. 12, pp. 2815-2819, Dec. 2021, doi: 10.1109/LWC.2021.3118269. 
\bibitem{b6}S. Wu, et al., "Blockage Prediction Using Wireless Signatures: Deep Learning Enables Real-World Demonstration," in IEEE Open Journal of the Communications Society, vol. 3, pp. 776-796, 2022, doi: 10.1109/OJCOMS.2022.3162591.
\bibitem{b7}Hasani, Ramin, et al. "Liquid Time-constant Networks." arXiv, 2020,  https://doi.org/10.48550/arXiv.2006.04439.
\bibitem{b8}Lechner, Mathias, et al. “Neural circuit policies enabling auditable autonomy.” Nature Machine Intelligence 2 (2020): 642-652.
\end{thebibliography}
\end{document}